\documentclass[11pt,twoside]{article}

\usepackage{asp2006}
\usepackage{graphicx}

\begin{document}

\title{Relaxing the Small Particle Approximation for Dust-grain opacities in Carbon-star Wind Models}   

\author{Lars Mattsson, and
Susanne H\"ofner}

\affil{$^1$Dept. of Astronomy \& Space Physics, Uppsala University, Sweden}

\begin{abstract} 
  We have computed wind models with time-dependent dust formation and grain-size dependent opacities, where (1) the problem is simplified by assuming a
  fixed dust-grain size, and where (2) the radiation pressure efficiency is approximated using grain sizes based on various means of the actual grain 
  size distribution. It is shown that in critical cases, the effect of grain sizes can be significant. For well-developed winds, however, the effects
  on the mass-loss rate and the wind speed are small.
\end{abstract}

{\bf Introduction.} In a new grid of C-star wind models Mattsson et al. (2010, A\&A, 509, 14) used the so-called {\it small-particle limit} (SPL) of Mie theory, i.e.,
grains are assumed to be small compared to the photon wavelengths when calculating their optical properties.
In that limit the wavelength- and grain size-dependence of the opacity can be greatly
simplified. Fig. 1 shows the distribution of actual grain sizes found in a representative subsample of models taken
from the model grid (presented in Mattsson et al. 2010, A\&A, 509, 14). The maximum grain size $a_{\rm d}$ (the "dividing radius" between the SPL and 
non-SPL regemes) where deviations in opacities from the small particle limit are less than 10\% is marked by the vertical dashed line.
The vast majority of the grid models did not meet this criterion (see Fig.~1).

{\bf Models.} We have re-computated a selection of models adopting different relaxations of the small-particle approximation (SPA) and picked models such
that two categories emerge: those with strong well-developed winds and those with slow critical winds (Mattsson \& H\"ofner 2010, A\&A, submitted).
We use an "optimised" (OPT) constant grain radius $a_{\rm gr} = 3.55\cdot 10^{-5}$ cm for all grains (the peak in $Q_{\rm rp}$, see Fig.1).
We also consider mean grain radii derived from one of the moments $K_i$ ($i = 1,2,3$) of the actual grain-size distribution as effective grain sizes used in the
raditive transfer calculations.

{\bf Results and Conclusions.}
In the critical-wind cases, the effect of grain sizes can be significant. Mass-loss rates may increase
by a factor of two, or more, and wind speeds by as much as an order of magnitude (see Fig.~2, open symbols).
Furthermore, the corresponding models with grain-size dependent opacities that have resultant winds tend to have much lower degrees of dust
condensation, compared to their SPA counterparts (again, see Fig.~2). Consequently, the "dust-loss rates" are much lower in these new models.
In contrast, for well-developed dust-driven winds (Fig.~2, filled symbols), where the dust formation has saturated, the effect of grain sizes
on mass-loss rate, wind speed and dust-to-gas ratio is almost negligible.

\acknowledgements
This work was partly supported by the Swedish Research Council 
(Vetenskapsr\aa det).

  \begin{figure}
  \begin{center}
  \includegraphics[width=35mm]{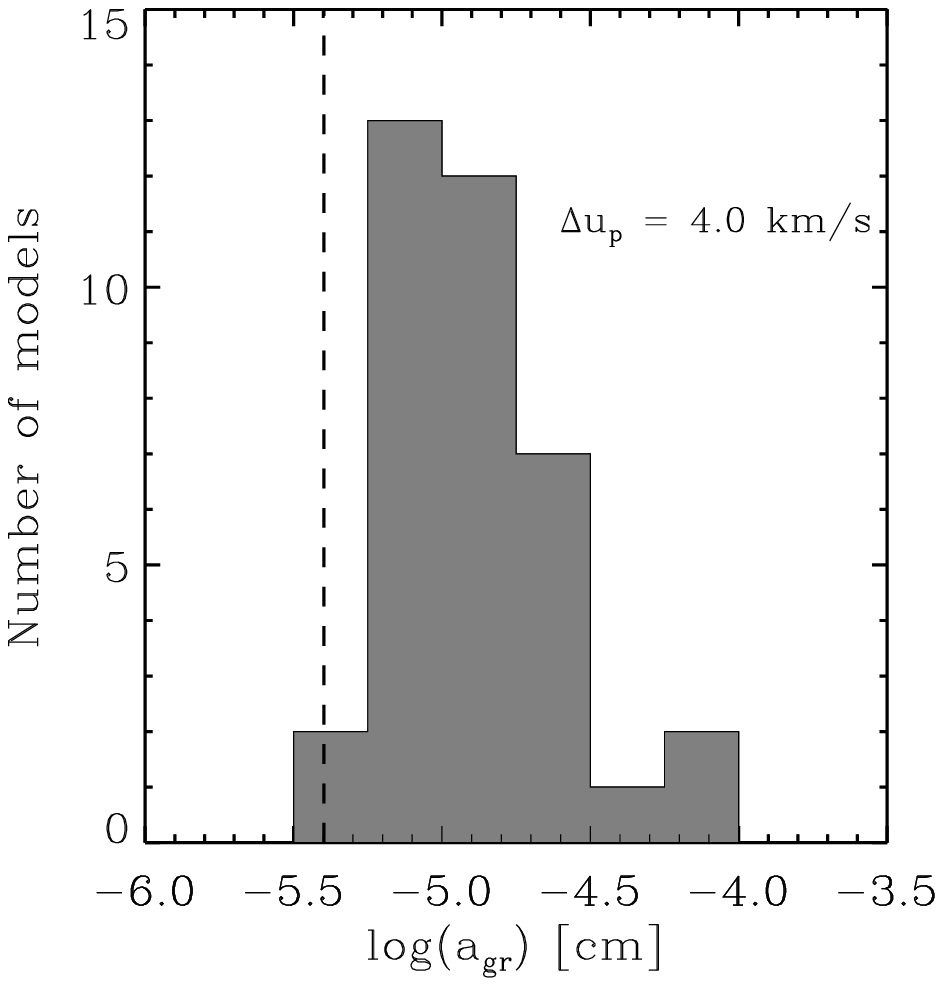}
  \includegraphics[width=35mm]{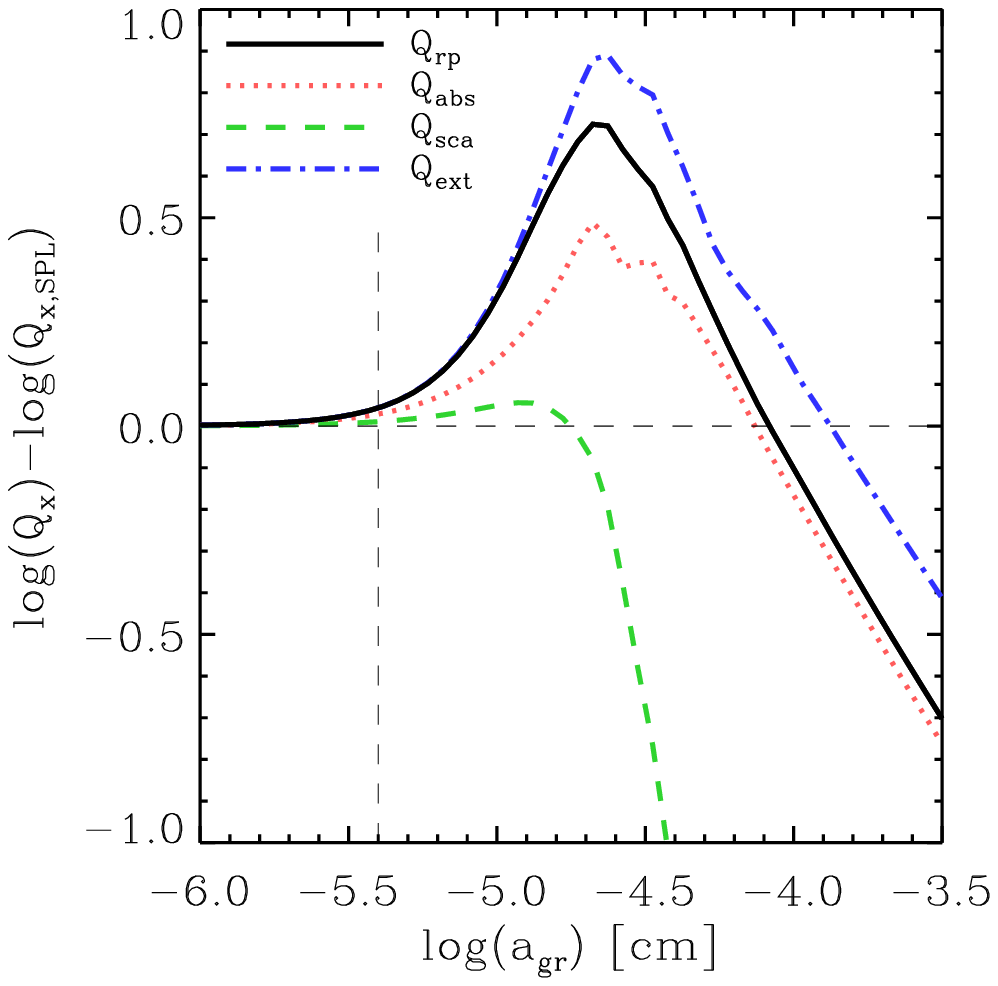}
  \end{center}
  \caption{Left: Histogram of grain sizes for wind-forming models with $M_\star=1M_\odot$, $\Delta u_{\rm p} = 4$ km s$^{-1}$ taken from Mattsson
  et al. (2010,  A\&A, 509, 14). The vertical dashed line marks the dividing radius $a_{\rm d}$. Right: The radiative pressure efficiency factor $Q_{\rm rp}$ 
  and its components $Q_{\rm ext}$, $Q_{\rm abs}$ and $Q_{\rm sca}$, relative to the corresponding SPL values used in the SPA models, as functions of grain
  radius at $\lambda=1\mu$m.
  }
  \end{figure}

 \begin{figure}
 \begin{center}
 \includegraphics[width=28mm]{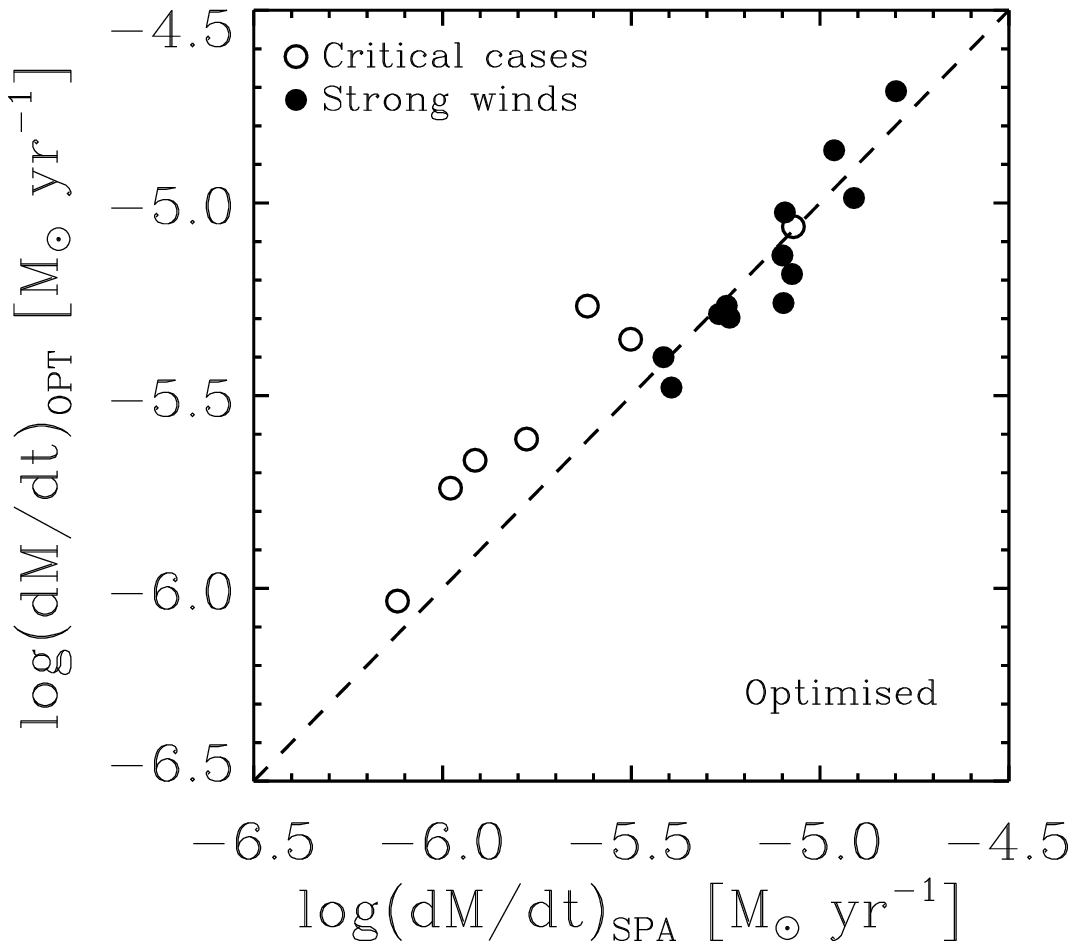}
 \includegraphics[width=28mm]{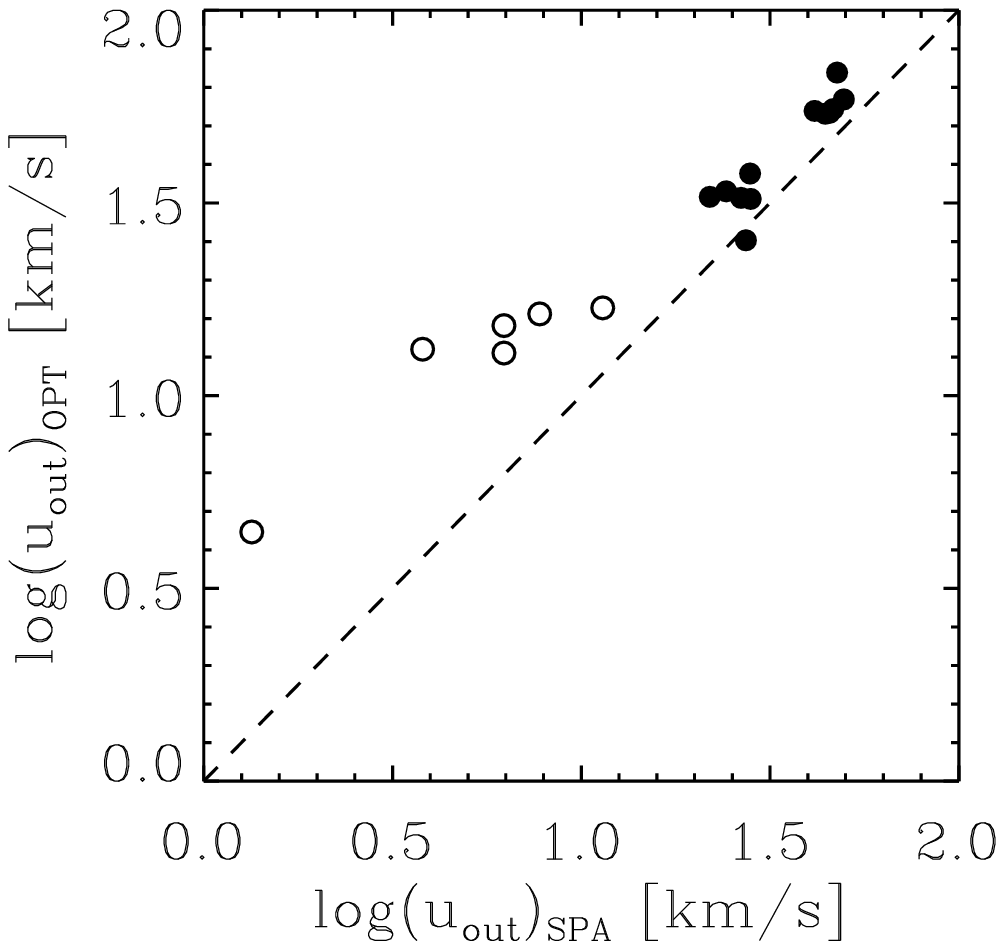}
 \includegraphics[width=28mm]{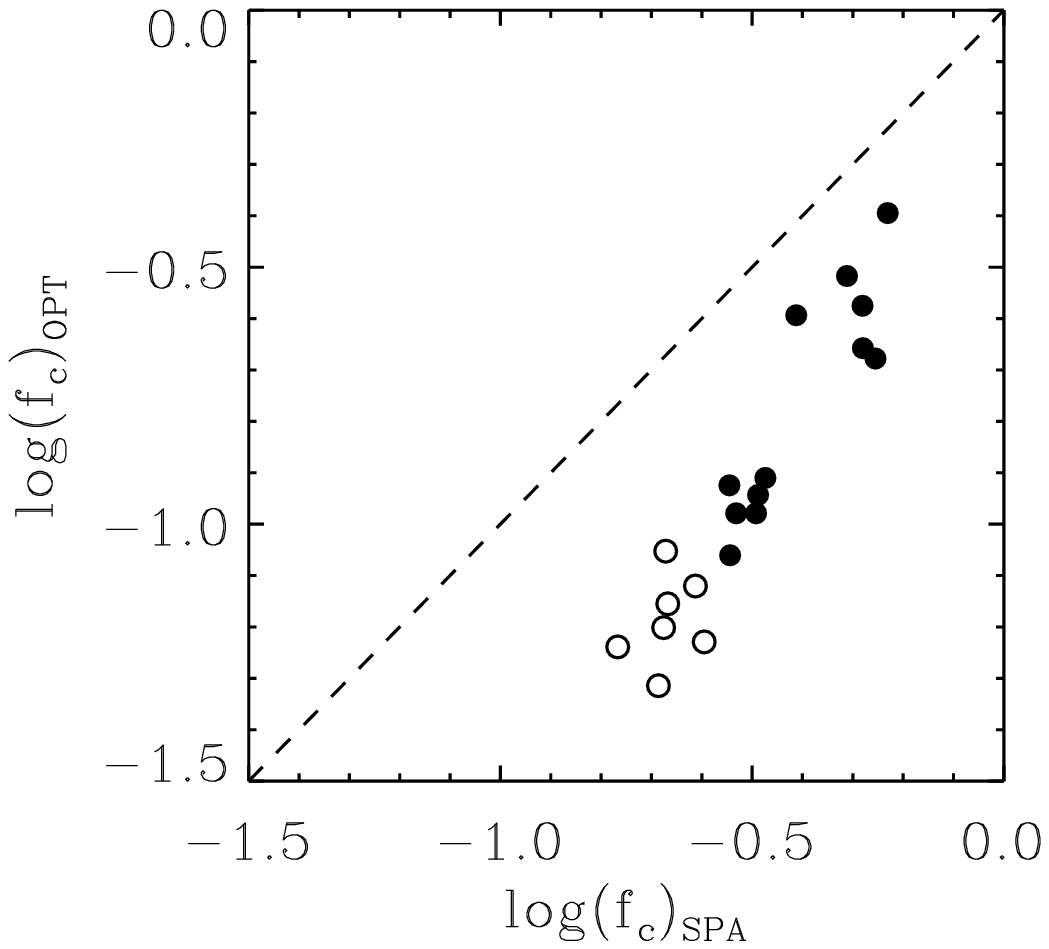}
 \includegraphics[width=28mm]{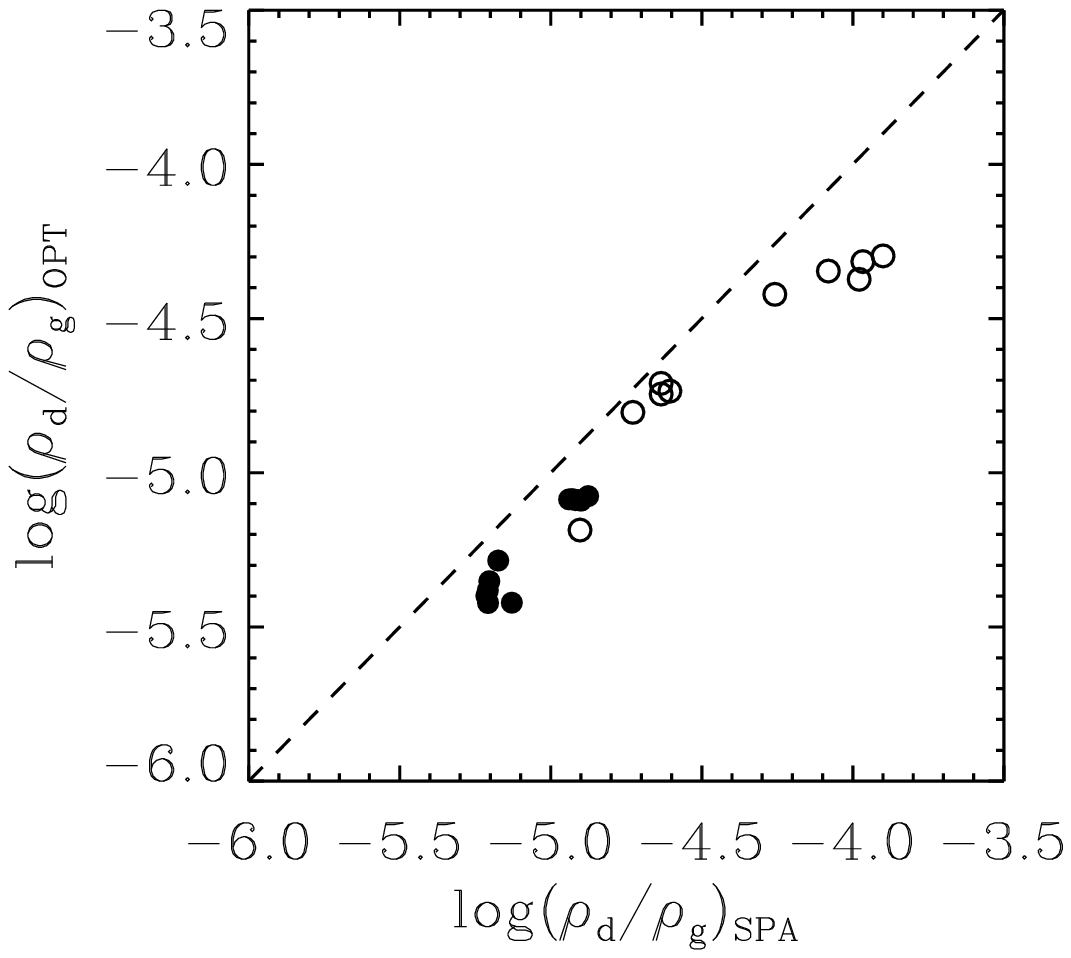}\\

 \includegraphics[width=28mm]{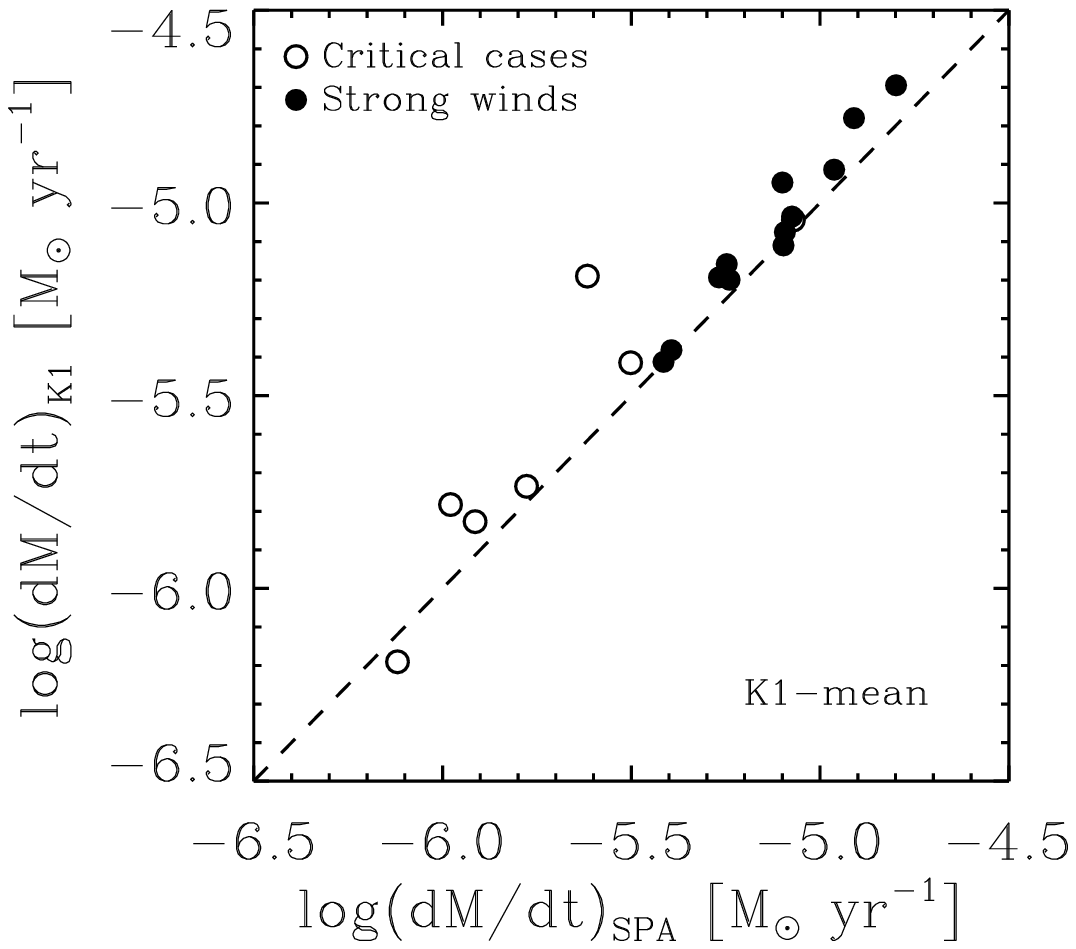}
 \includegraphics[width=28mm]{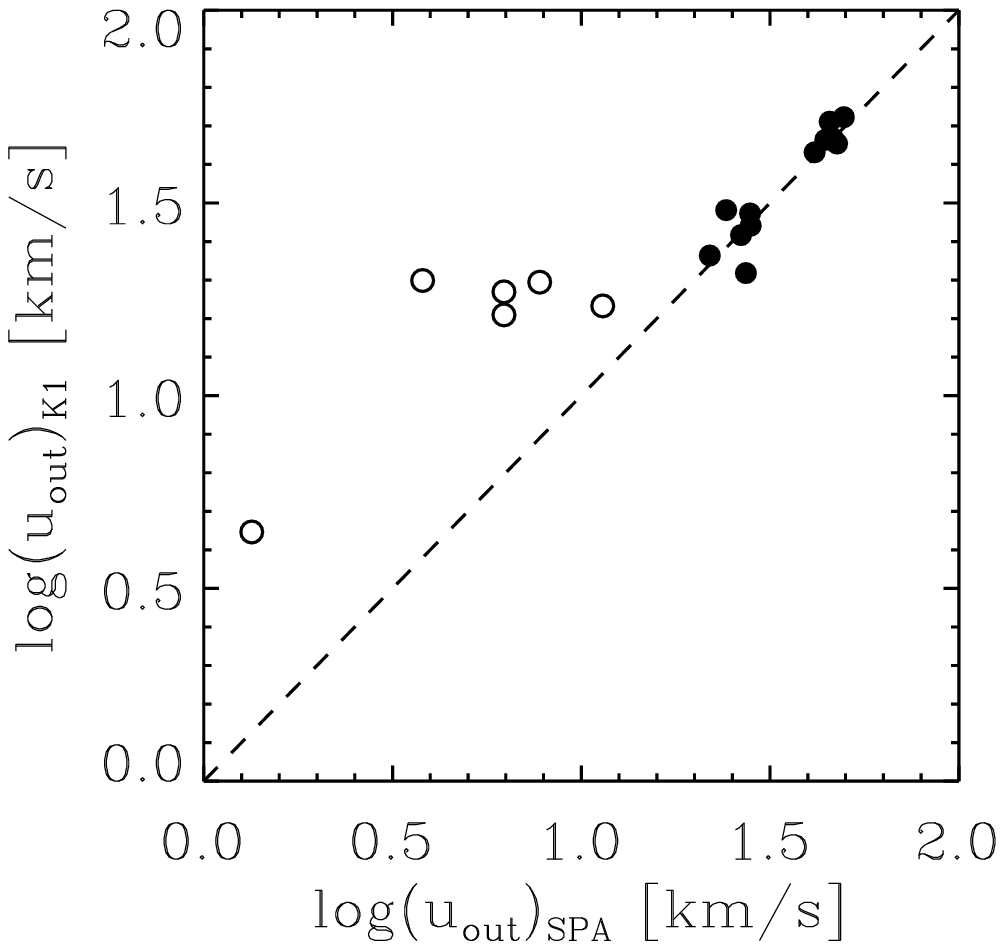}
 \includegraphics[width=28mm]{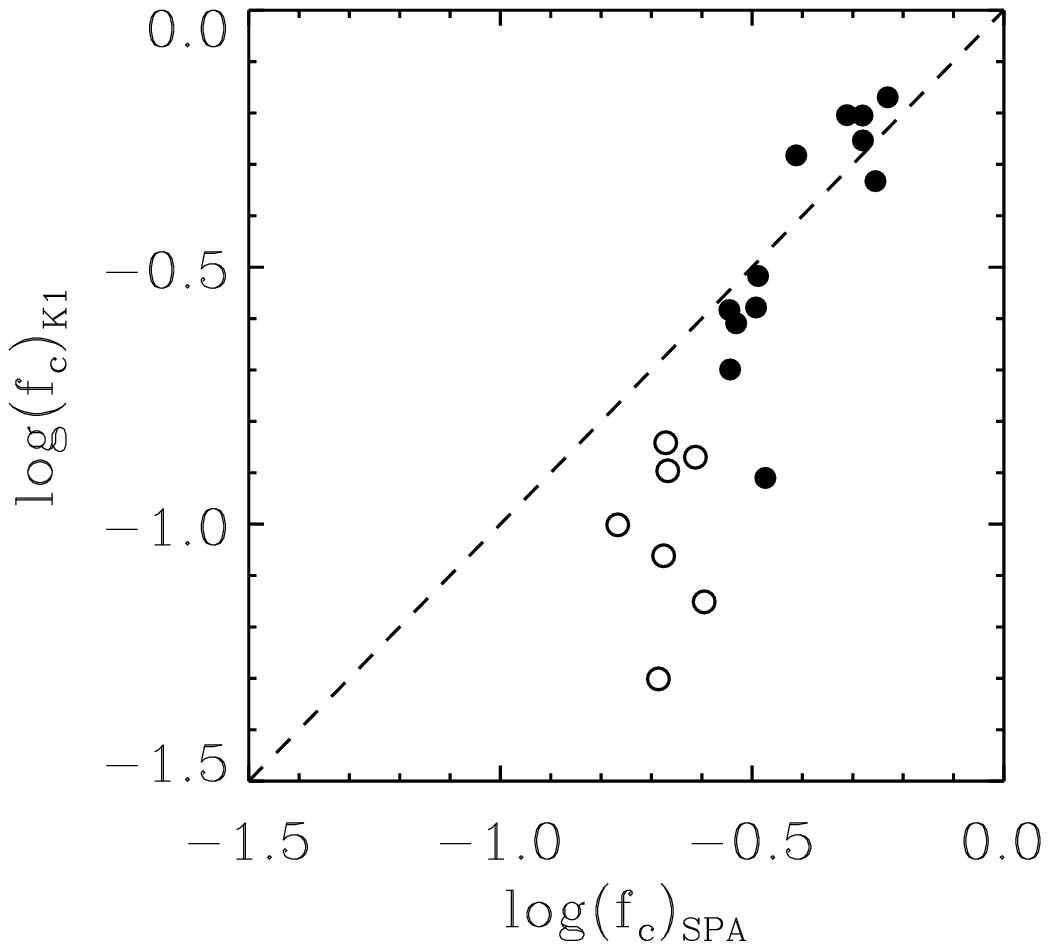}
 \includegraphics[width=28mm]{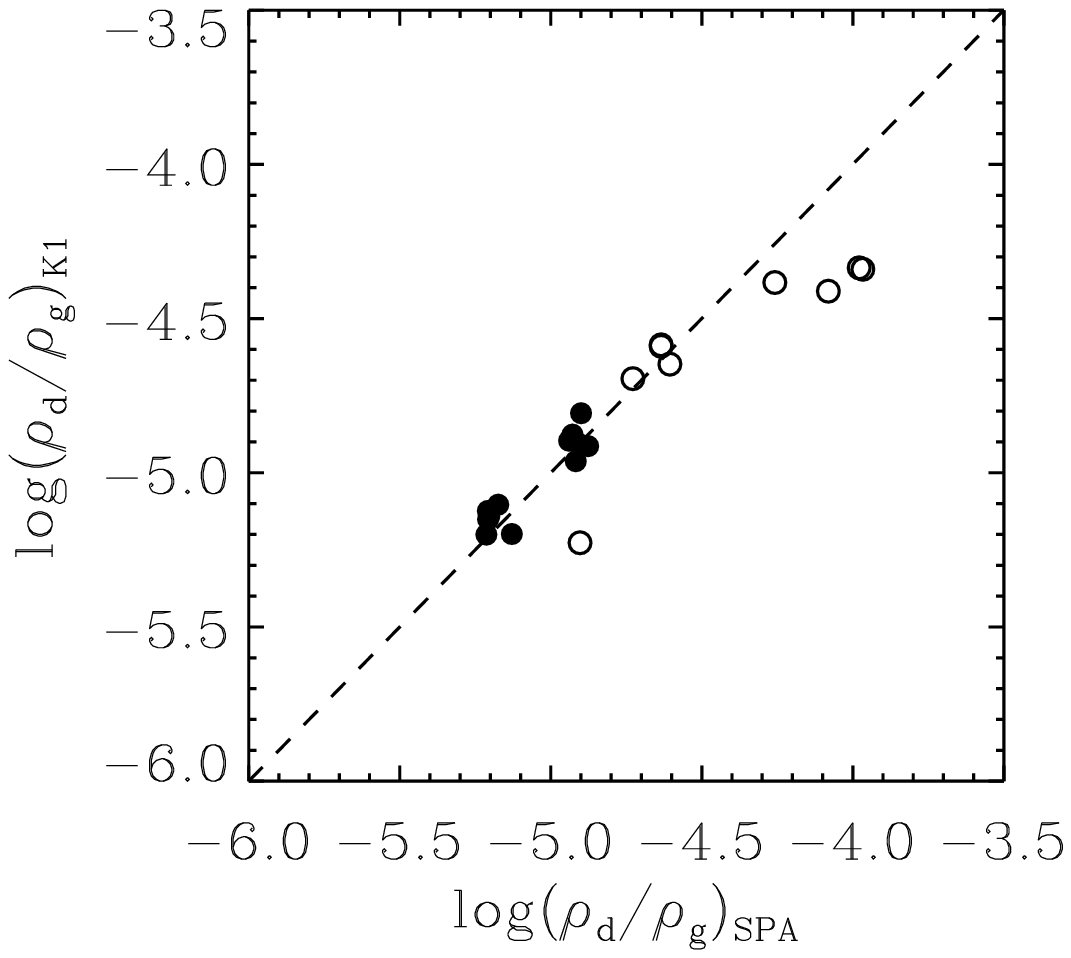}\\

 \includegraphics[width=28mm]{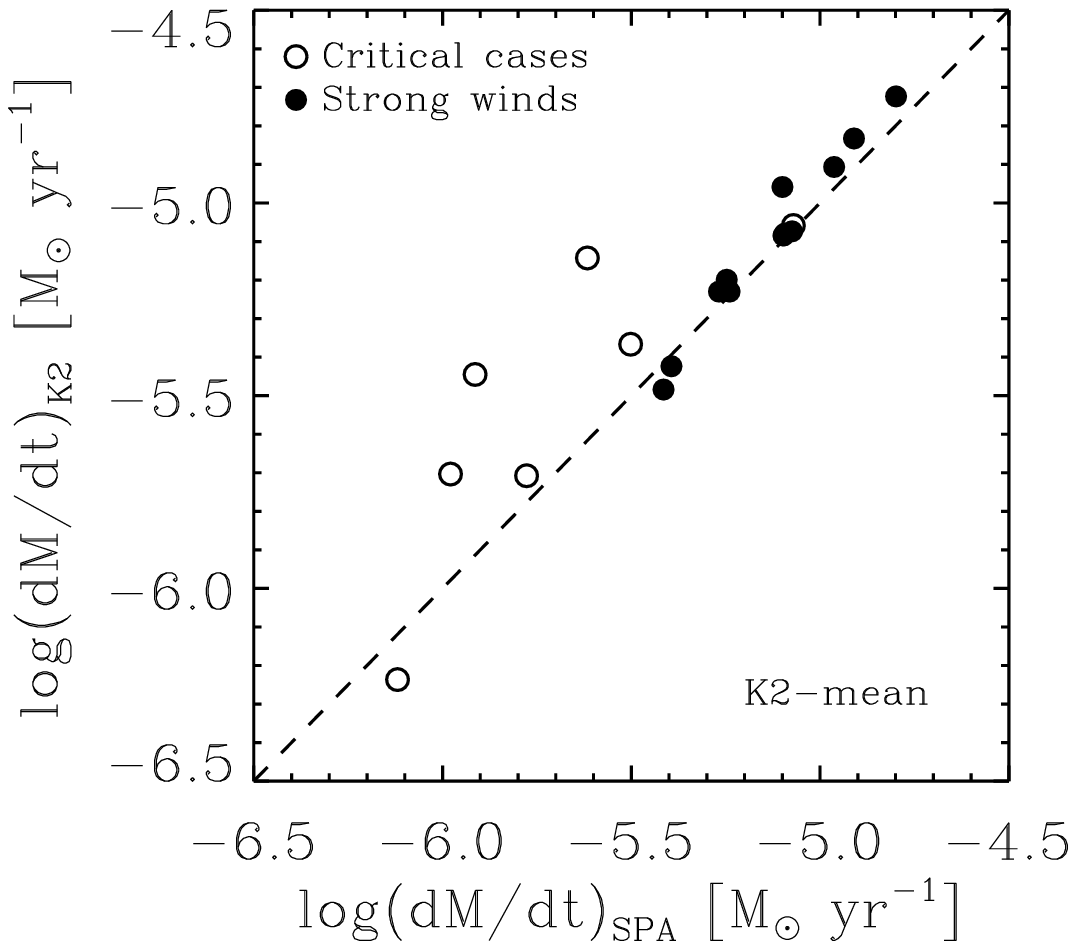}
 \includegraphics[width=28mm]{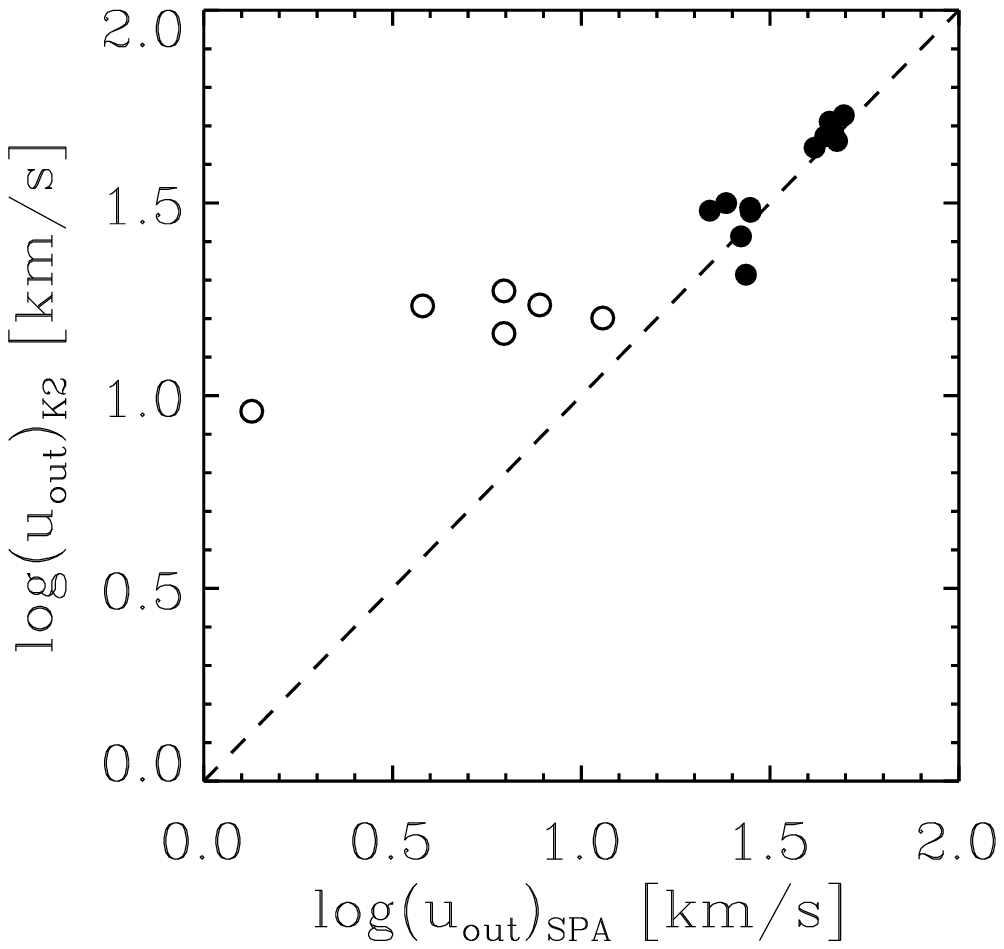}
 \includegraphics[width=28mm]{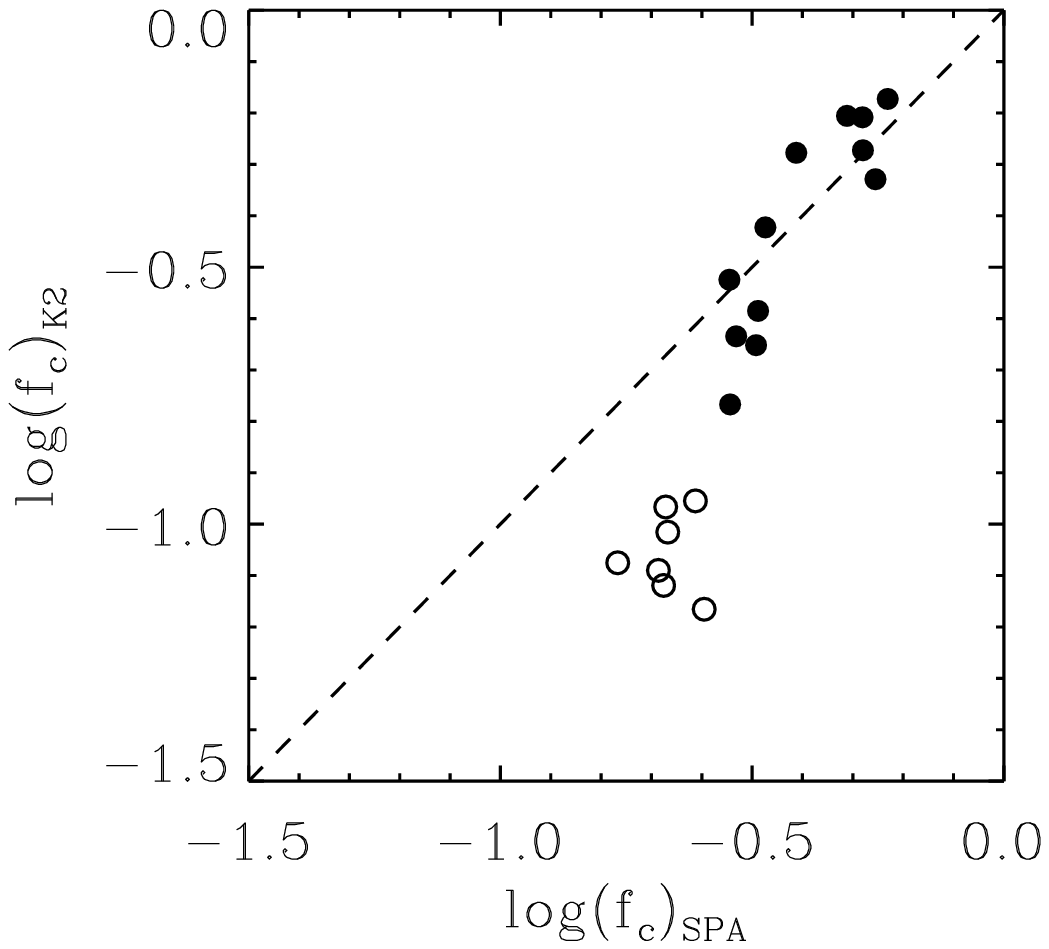}
 \includegraphics[width=28mm]{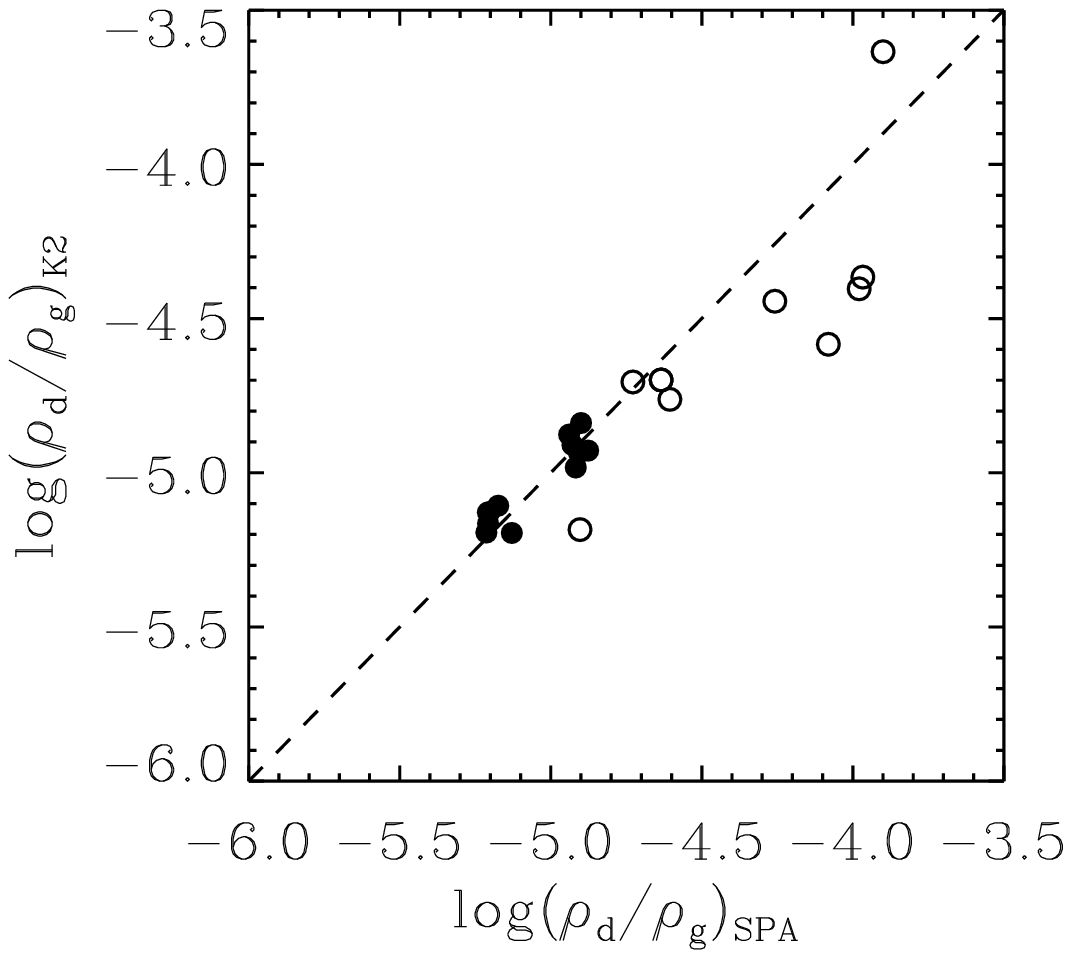}\\

 \includegraphics[width=28mm]{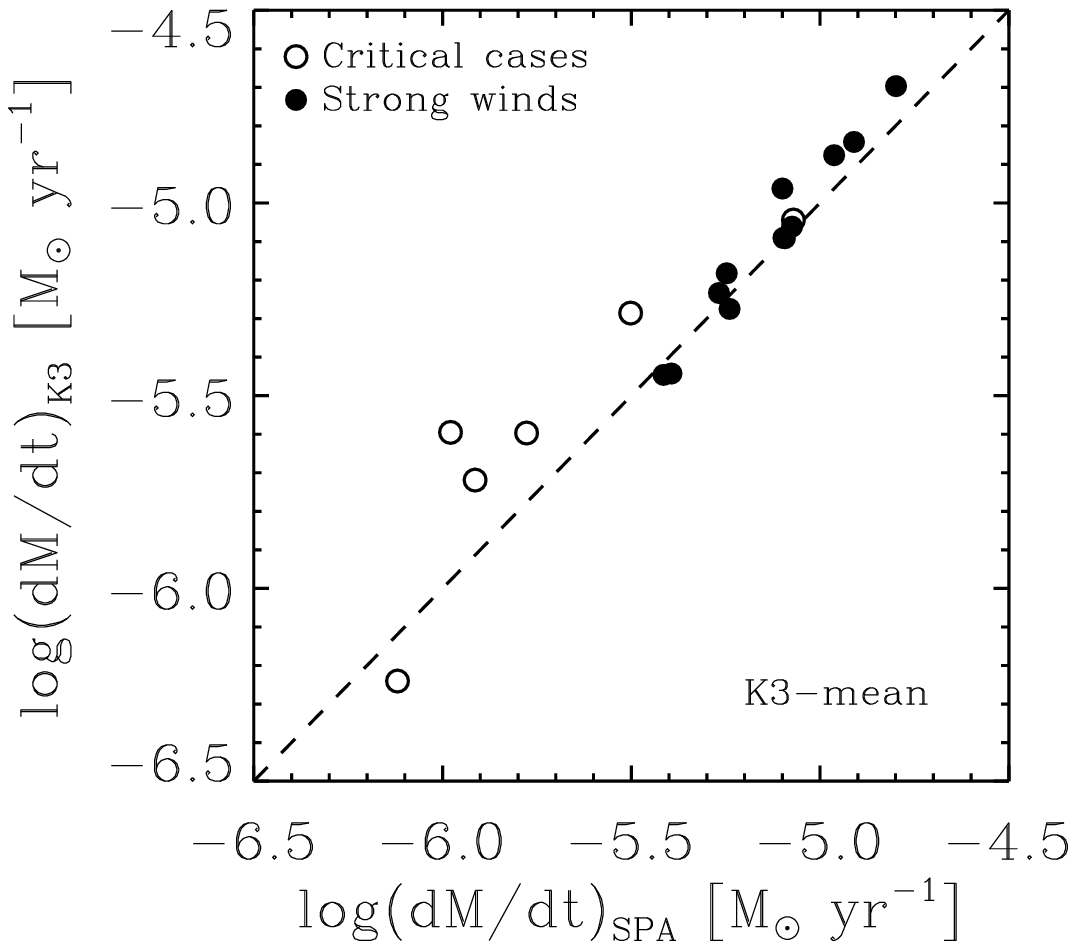}
 \includegraphics[width=28mm]{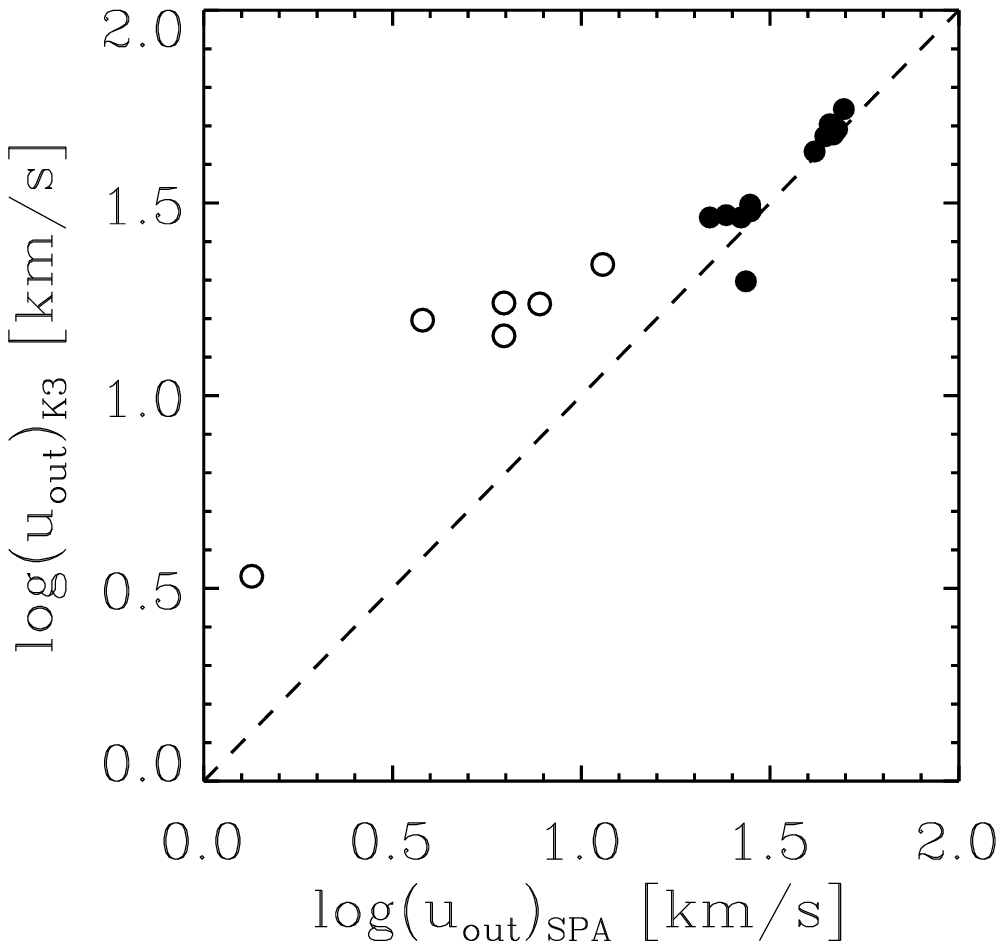}
 \includegraphics[width=28mm]{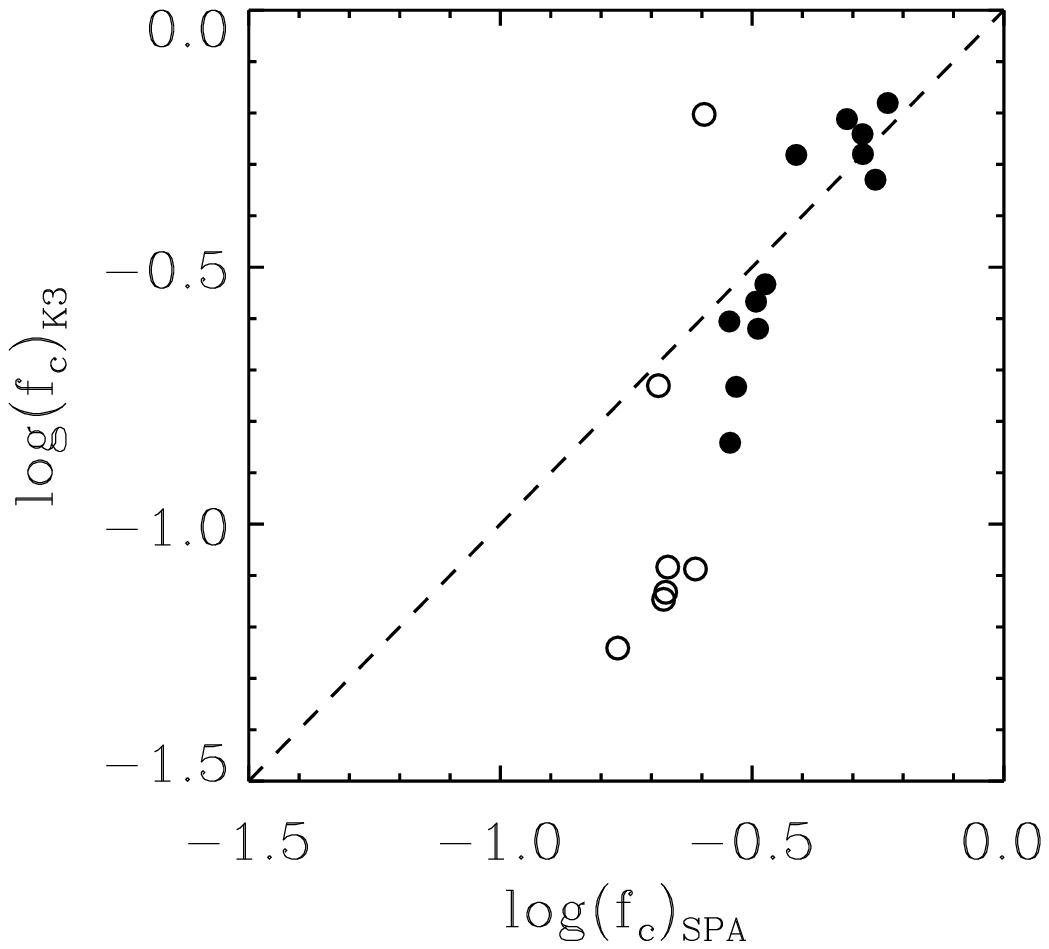}
 \includegraphics[width=28mm]{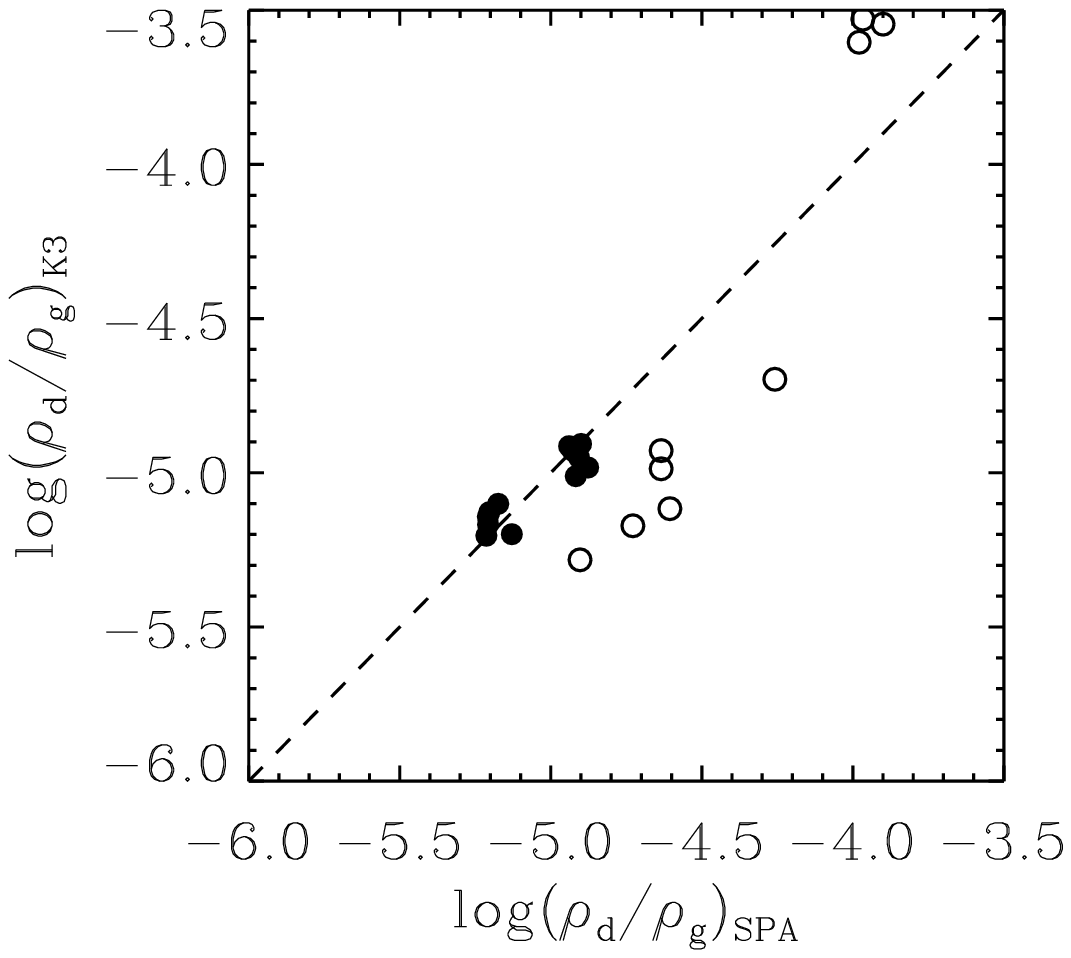}
 \end{center}
 \caption{From left to right: Mass-loss rates, wind speeds, mean degrees of dust condensation and dust-to-gas ratios for
 "$Q_{\rm rp}$-optimised" and K1, K2, K3 models versus the corresponding SPA models.  The dashed lines show the case of equal values.
 \label{agr_spa_Kopt}
  }
  \end{figure}

\end{document}